\documentclass[epj,final]{svjour}
\usepackage{graphicx}

\begin{document}

\title{Thermodynamic consistency for  nuclear matter 
calculations}

\author{P. Bo\.{z}ek 
 \and P. Czerski  
}
\institute{  Institute of Nuclear Physics, PL-31-342 Krak\'{o}w, Poland }
\date{\today}

\abstract{
We investigate the relation between the binding energy 
and the Fermi energy and between different expressions for the 
pressure in cold nuclear matter. 
For a self-consistent
calculation based on a $\Phi$ derivable $T-$matrix approximation with 
off-shell propagators the thermodynamic relations are well satisfied unlike
for a $G-$matrix or a $T-$matrix approach using quasi-particle propagators 
in the ladder diagrams.
}


\maketitle


\section{Introduction}

Nuclear matter calculations are usually performed using Brueckner
type resummation of ladder diagrams. Works using realistic interactions
lead to reasonable results for the saturation density and the 
binding energy at the 
saturation point. However in violation of the Hugenholz-Van Hove theorem
the resulting Fermi energy $E_F$
 at the saturation point is usually very different from 
the binding energy per particle $E/N$. It is a manifestation of a general 
violation of thermodynamic consistency by the $G-$matrix approximation.
The problem was discussed in the literature \cite{gmhvh,jong}
 and improvements due
to rearrangement terms were invoked but without removing the discrepancy 
altogether. Improvement of the fulfillment of the Hugenholz-Van Hove
property
with respect to the $G-$matrix approximation is observed
when using the quasi-particle 
$T-$matrix  approach, or correction from hole-hole lines 
\cite{gmhvh,jong}.

On the other hand it is known that the exact theory \cite{hvh,lw,baym} should 
fulfill certain thermodynamical relations. The simplest one being the 
exact equality of the Fermi momenta for the free and the interacting theory.
Another statement that we shall consider in the present work is the 
equivalence of two ways of calculating the pressure in a system at zero 
temperature~:
\begin{eqnarray}
\label{press}
\label{press1}
P&=& \rho^2 \frac{\partial (E/N)}{\partial \rho} \\
\label{press2}
& =& \rho (E_F - E/N) \ ,
\end{eqnarray}
where $\rho$ is the nuclear matter density.
From the above relation follows that at the saturation point where 
$(E/N)$ has a minimum
\begin{equation}
E_F=E/N \ ,
\end{equation}
i.e. the Hugenholz-Van Hove property. 
These relations are satisfied by the exact theory and 
 can also  be satisfied  in a perturbative calculation to a given 
order of the expansion parameter.

\begin{figure}

\centering
\includegraphics[width=0.45\textwidth]{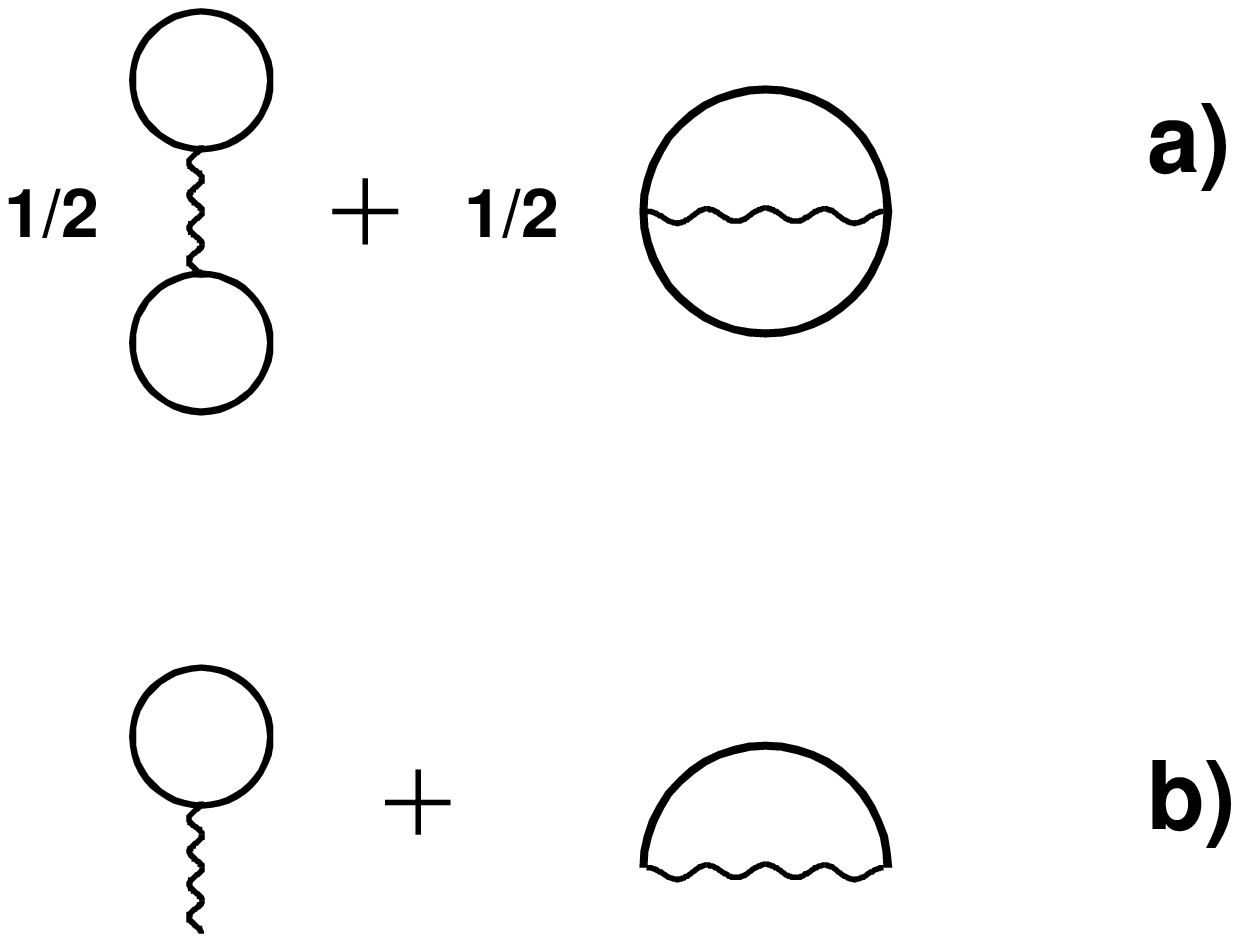}
\includegraphics[width=0.45\textwidth]{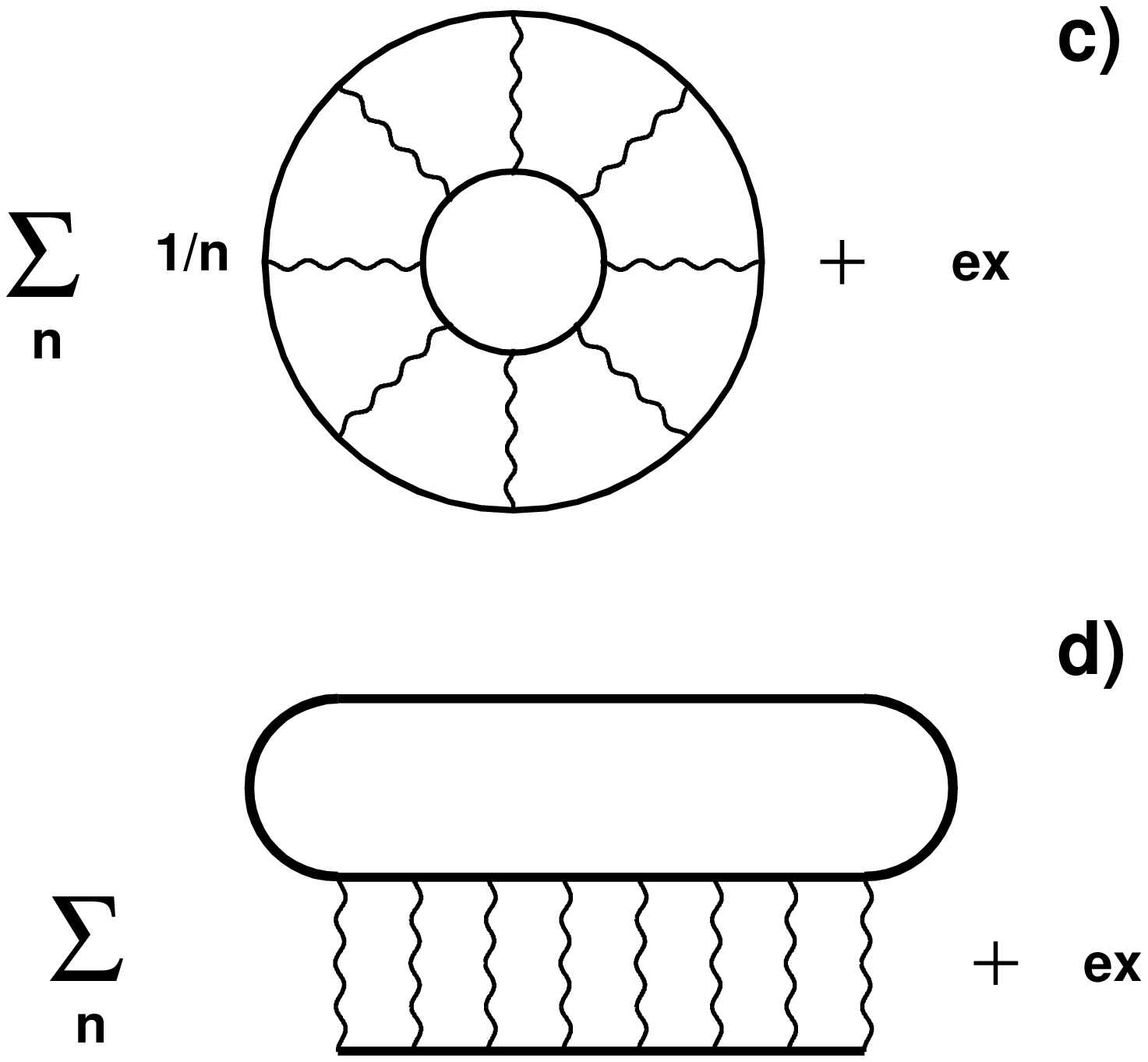}
\caption{Diagrams contributing to the generating functional $\Phi$ in
the Hartree-Fock approximation {\bf a)} and in the $T-$matrix approximation 
{\bf c)}. The corresponding diagrams for the self-energy are shown in parts
{\bf b)} and {\bf d)} respectively.}
\label{phihf}
\end{figure}

Non-perturbative approximations schemes which are ther\-mo\-dynamically 
consistent
are know \cite{baym}. 
Baym has shown that the condition of the thermodynamical consistency of an
 approximation can be related to the so called $\Phi$ derivability.
The self-energy is constructed as a functional derivative of a functional
$\Phi$ of dressed propagators $G(k)$ and bare vertices
\begin{equation}
\label{sigphi}
\Sigma(k)=\frac{\delta \Phi}{\delta G(k)} \ .
\end{equation}
The approximate functional $\Phi$ is defined by
 a set of two-particle irreducible   diagrams. 
$\Phi$ derivable approximations to the self-energy are also termed as
 conserving approximations since they lead to conservation 
laws in corresponding
transport equations \cite{kb}.
In particular different types of non-perturbative approximations can be 
identified for the generating functional. Below we shall consider two of them
the Hartree-Fock approximation and the $T-$matrix
approximation  (Fig. \ref{phihf}).
 Diagrams for the corresponding self-energies obtained by taking a 
functional derivative are also shown in Fig. \ref{phihf}.
 It must be stressed again that the propagators  
in the diagrams for $\Phi$ are dressed self-consistently by the self-energy
(\ref{sigphi}). For the Hartree-Fock approximation it means only a shift 
in the single-particle energies, but for the $T-$matrix approach
one has to take into account the full spectral function for the
propagators in the calculation of $\Phi$
or $\Sigma$.
Calculations involving off-shell propagators in the $T-$matrix ladder
have been recently performed \cite{dickhoff,ja,ja2,gent} both in the
 normal and in the superfluid phase. Below we shall restrict ourselves to 
zero temperature normal nuclear matter.

\section{Approximations for the nuclear matter problem}

We shall compare different calculation of cold nuclear matter with a model 
interaction. We choose a separable rank two parameterization of Mongan type
\cite{mongan} in
 the $S$ wave with softened repulsive core
\begin{equation}
V_\alpha(k,p)=\lambda^r_\alpha g^r_\alpha(k)g^r_\alpha(p)
-\lambda^a_\alpha g^a_\alpha(k)g^a_\alpha(p)
\end{equation}
with $g^{r,a}(p)=\frac{1}{p^2+\beta_{r,a}^2}$
and
\begin{eqnarray}
\label{monganpar}
\lambda^r=29.6 GeV^2 & \beta^r=639 MeV & \nonumber \\
\lambda^a=2.91 GeV^2 & \beta^a=352 MeV & \ \ {\rm for } \ \ \
\alpha=S^1_0 \nonumber \\
\lambda^r=5.27 GeV^2 & \beta^r=471 MeV & \nonumber \\
\lambda^a=4.78 GeV^2 & \beta^a=376 MeV & \ \ {\rm for } \ \ \
\alpha=S^3_1 \ \ .
\end{eqnarray}

With this interaction nuclear matter properties will be calculated
 within the following approximations

\begin{itemize}
\item
Brueckner 
resummation of particle-particle ladder 
diagrams with in medium $G-$matrix 
\begin{eqnarray}
\label{gmatrix}
& &<{\bf p}|G
({\bf P},\Omega)|{\bf p}^{'}> = V
({\bf p},{\bf p}^{'}) \nonumber \\& &
 + 
\int\frac{d^3q}{(2 \pi)^3} V
({\bf p},{\bf q}) 
\frac{(1-f(\omega_{p_1}))(1-f(\omega_{p_2}))}
{\Omega-\omega_{p_1}-\omega_{p_2}} \nonumber \\ & & 
 <{\bf q}|G({\bf P},\Omega)
|{\bf p}^{'}> \ ,
\end{eqnarray}
where ${\bf p_{1,2}}={\bf P}/2\pm {\bf q}$.
$G-$matrix resummation  allows to define 
single particle energies and gives  relatively good results
for the saturation properties of nuclear matter.
In the above equation and in the following we skip the spin, isospin indices
which are implicitly summed over.
Medium effects enter through the Pauli blocking factors 
$1-f(\omega_p)$ in the numerator and single-particle energies $\omega_{p}$ 
in the denominator. 
 The single particle energies $\omega_{p}$, are
self-consistently defined by the G-matrix 
\begin{equation}
\omega_p=\frac{p^2}{2 m}+U(p,\omega_p)
\end{equation}
where
\begin{eqnarray}
& & U(p,\omega)=\int \frac{d^3k}{(2 \pi)^3} f(\omega_k)\nonumber \\ & &
<({\bf p-k})/2| G(|{\bf p+k}|,\omega_k+\omega)|({\bf p-k})/2>
 \ .
\end{eqnarray}

\item
 In the quasi-particle $T-$matrix
approximation \cite{vo,roepke} the ladder diagrams include 
both particle-particle and hole-hole propagation. The Pauli blocking factor 
$(1-f(\omega_{p_1}))(1-f(\omega_{p_2}))$ in the G-matrix equation 
is replaced 
by  $1-f(\omega_{p_1})-f(\omega_{p_2})$ in the equation for
the retarded T-matrix
\begin{eqnarray}
\label{tmatrixqp}
& & <{\bf p}|T
({\bf P},\Omega)|{\bf p}^{'}>  =  V
({\bf p},{\bf p}^{'})  + \nonumber \\ & & 
\int\frac{d^3q}{(2 \pi)^3} V
({\bf p},{\bf q}) 
\frac{(1-f(\omega_{p_1})-f(\omega_{p_2}))}
{\Omega-\omega_{p_1}-\omega_{p_2}+i\epsilon} \nonumber \\ & & 
 <{\bf q}|T({\bf P},\Omega)
|{\bf p}^{'}> \ .
\end{eqnarray}
The  imaginary part of the retarded 
self-energy in the T-matrix approximation is 
\begin{eqnarray} & &
{\rm Im} \Sigma(p,\omega)=\int \frac{d^3k}{(2 \pi)^3} 
\Big(f(\omega_k)+b(\omega+\omega_k)\Big) \nonumber \\ & &  
<({\bf p-k})/2| {\rm Im} T(|{\bf p+k}|,\omega_k+\omega)|({\bf p-k})/2>
\ ,
\end{eqnarray}
where $b(\omega)$ is the Bose distribution. 
The real part of the self energy  consists of the Hartree Fock self-energy 
and a dispersive contribution obtained from ${\rm Im}\Sigma$
\begin{equation}
\label{disper}
{\rm Re}\Sigma(p,\omega)=\Sigma_{HF}(p)+{\cal P}\int\frac{d\omega^{'}}{\pi}
\frac{{\rm Im}\Sigma(p,\omega^{'})}{\omega^{'}-\omega} \ .
\end{equation}
The imaginary part of the self-energy is  neglected leading to the 
quasi-particle approximation for the two-nucleon propagator in the 
T-matrix (Eq. \ref{tmatrixqp}).

\item

Allowing for off-shell propagation of  nucleons and taking the self-energy
self-consistently (also its imaginary part)
 requires the use of full spectral functions in the 
calculation resulting in   more complicated expressions for the $T-$matrix and
 the self-energy \cite{dickhoff,ja}
\begin{eqnarray}
\label{teq}
& & <{\bf p}|T({\bf P},\Omega)|{\bf p}^{'}> = V({\bf p},{\bf p}^{'})
\nonumber \\ & & + 
 \int\frac{d\omega_1}{2\pi}\int\frac{d\omega_2}{2\pi}
\int\frac{d^3q}{(2 \pi)^3} V({\bf p},{\bf q})
\frac{\big(1-f(\omega_1)-f(\omega_2)\big)}
{\Omega-\omega_1-\omega_2+i\epsilon} \nonumber \\ & &
A(p_1,\omega_{1})A(p_2,\omega_{2})
 <{\bf q}|T({\bf P},\Omega)
|{\bf p}^{'}> 
\end{eqnarray}
and
\begin{eqnarray}
\label{ims}& & 
{\rm Im}
\Sigma(p,\omega) =\int\frac{d\omega_1}{2 \pi}\int \frac{d^3k}{(2 \pi)^3}
A(k,\omega_1) \nonumber \\ 
& & <({\bf p}-{\bf k})/2|{\rm Im}T({\bf p}
+{\bf k},\omega+\omega_1)|({\bf p}-{\bf k})/2> \nonumber \\ & & 
 \Big( f(\omega_1)+b(\omega+\omega_1) \Big) \ ,
\end{eqnarray}
where 
\begin{equation}
A(p,\omega)=\frac{-2 {\rm Im}\Sigma(p,\omega)}{\Big(\omega-p^2/2m 
-{\rm Re}\Sigma(p,\omega)\Big)^2 +{\rm Im}\Sigma(p,\omega)^2}
\end{equation}
is the self-consistent spectral function of the nucleon.

\item
Finally we present results for a  simple Hartree-Fock approximation. It is
certainly not well suited for realistic applications in nuclear matter.
However this approach is $\Phi$ derivable and it is illustrative 
to check its thermodynamic consistency explicitly. The Hartree-Fock
approximation with parameters given by Eq. (\ref{monganpar}) shows 
no saturation. We reduced the repulsive part of the interaction
 $\lambda^r_\alpha$ by $1.15$ for the Hartree-Fock
calculation. This rescaling mimics the effect of ladder resummation 
which leads to a reduction of the repulsive core.
\end{itemize}
\noindent Equations for  all the approximations schemes have to be
solved iteratively, with a constraint on the total density.
The numerical method for the solution of the $T-$matrix
equation with off-shell propagators \cite{ja2} has been generalized to
the case of low and zero temperature. The details of the
numerical  procedure will be
given elsewhere.

\section{Results for thermodynamic properties around the saturation point}

\begin{figure}
\centering
\includegraphics[width=0.48\textwidth]{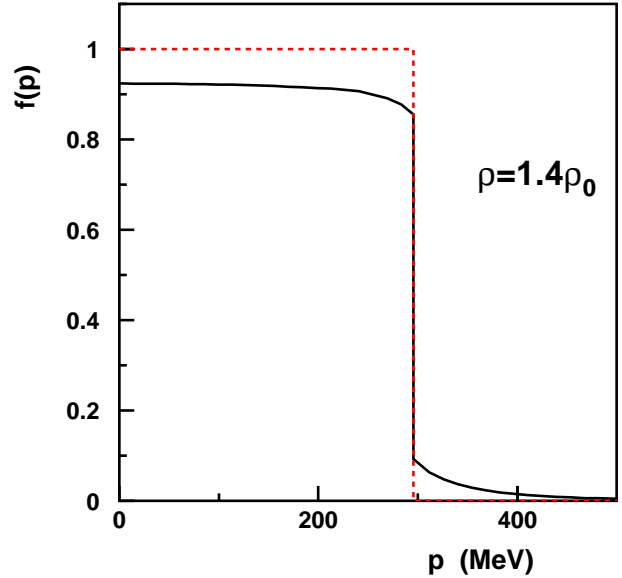}
\caption{Momentum distribution of nucleons for the full $T-$matrix 
calculation
compared to the free fermion distribution.}
\label{fermionfig}
\end{figure}

Only within the self-consistent $T-$matrix calculation is the
momentum distribution of nucleons 
\begin{equation}
n(p)=\int_{-\infty}^\mu \frac{d\omega}{2 \pi} A(p,\omega)
\end{equation}
different from the Fermi-Dirac distribution (Fig. \ref{fermionfig}).
\begin{figure}
\centering
\includegraphics[width=0.48\textwidth]{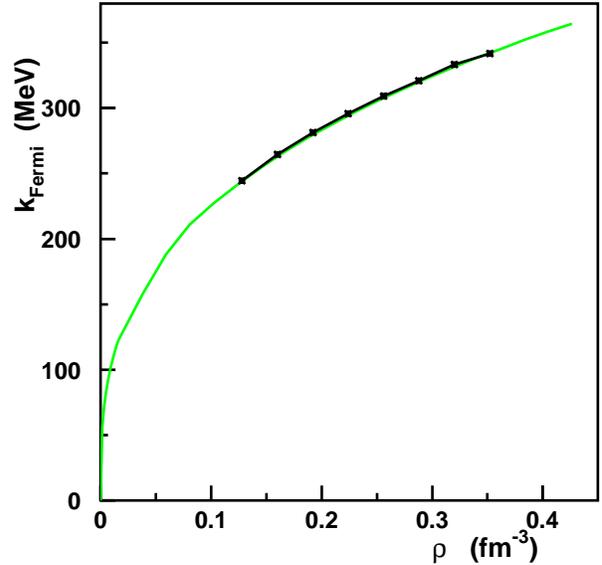}
\caption{Fermi momentum as obtained in the full $T-$matrix calculation (points)
compared to the Fermi momentum of the free fermion gas (solid line).}
\label{pfermifig}
\end{figure}
Clearly a Fermi liquid behavior is observed in the $T-$matrix approximation, 
with a jump in the fermion density of \\ $\bigg(1-\frac{\partial{\rm Re}
\Sigma(p_F,\omega)}{\partial \omega}|_{\omega=E_F}\bigg)^{-1}
\simeq .7$ at the Fermi momentum.
In the calculation the chemical potential $\mu=E_F$ is fixed by the constraint
on the total density
\begin{equation}
\int_{-\infty}^{\mu}
 \frac{d\omega}{2\pi}\int \frac{d^3p}{(2\pi)^3} A(p,\omega)=\rho \ .
\end{equation}
The corresponding Fermi momentum $p_F$ is defined by $E_F=\omega_{p_F}$~.
For a conserving approximation the Fermi  momentum should be the same as
 the Fermi
momentum of a free fermion gas \cite{lutt2,baym}.
 Indeed it is well satisfied for a range
of densities for the self-consistent $T-$matrix calculation
(Fig \ref{pfermifig}). All the other approximation discussed in this work
 fulfill 
this relation trivially 
since they use quasi-particles.

The energy per particle in the different approximations can be obtained from
 the standard form of the energy density 
\begin{equation}
E/N=\frac{1}{\rho}
\int\frac{d^3p}{(2 \pi)^3}\frac{d\omega}{2\pi}
\frac{1}{2}\bigg(\frac{p^2}{2 m}+
\omega \bigg) A(p,\omega)f(\omega) \ .
\end{equation}
Only for the self-consistent $T-$matrix the spectral function and the 
$\omega$ integration is nontrivial. 
For the other approximation schemes the spectral function is a delta function.
In that case
 the energy per particle can be expressed in the usual way using the
single particle potential and kinetic energies.

\begin{figure}

\centering
\includegraphics[width=0.48\textwidth]{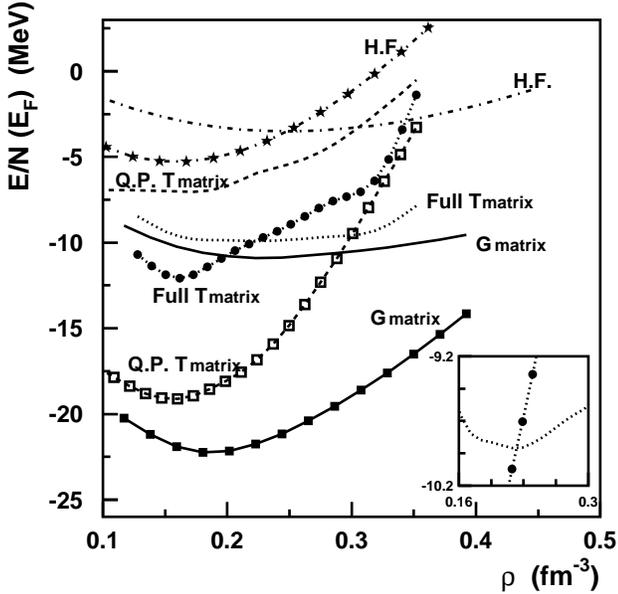}
\caption{Binding energy per particle for the 
$G-$matrix calculation (solid line), the 
on shell $T-$matrix calculation (dashed line), the full $T-$matrix calculation
(dotted line) and the Hartree-Fock calculation (dashed-dotted line)
as function of the  density.
The corresponding Fermi energies
 are denoted by the same lines with solid boxes, open boxes,
full circles and stars for the $G-$matrix, on shell $T-$matrix, full $T-$matrix
and Hartree-Fock results. In the insert is show a blow up of the region around
the saturation point for the full $T-$matrix calculation.}
\label{enfig}
\end{figure}

In Fig. \ref{enfig} is plotted the energy per particle for different 
approximations for a range of densities aground the saturation density.
The $G-$matrix and the full $T-$matrix calculations give very similar 
results for the binding energy. The $T-$matrix with 
quasi-particle propagators
gives somewhat different results, with lower saturation density and smaller 
binding energy. This behavior is due to very strong modifications of 
the effective mass around the Fermi momentum in the 
quasi-particle $T-$matrix approach. This effect is caused by the appearance 
of the pairing singularity in the $T-$matrix \cite{roepke,schnell,ja2}.
In fact the quasi-particle $T-$matrix approximation is oversensitive to the
presence of the pairing singularity, 
since the use of full spectral functions 
reduces the influence of the Cooper pair bound state on the nucleon 
spectral function and the single particle energies \cite{ja2,ja3}.

In the Table are shown the corresponding binding energies and 
saturation densities.  
\begin{table}
\begin{tabular}{|r|c|c|c|l|}
\hline
approximation & $\rho_s/\rho_0$ & $E_F$  & $E/N$ & ${\cal K}$ 
\\
& & (MeV) & (MeV) & (MeV) \\
\hline
Hartree-Fock & 1.55 & -3.5 & -3.5 & 87 \\& & & & \\
$G-$matrix & 1.42 & -21.6& -10.9 & 107 \\& & & & \\
$T-$matrix on shell &1.08  &-18.9 &-7.0 &103 \\  & & & & \\
 $T-$matrix& 1.39 &-9.9 &-9.9 & 103\\ 
\hline
\end{tabular}
\caption{Saturation density, Fermi energy, binding energy and 
compression modulus for different approximations discussed in the text}
\end{table}
 The Hartree-Fock approximation 
gives significantly different results, only after a change in the parameters 
it has a saturation point at all. 
The Fermi energy obtained for 
different densities depends very much on the approximation chosen.
Only for consistent approaches, i.e. Hartree-Fock and self-consistent 
$T-$matrix is the Hugenholz- Van Hove condition at the saturation point 
satisfied. The difference between $E_F$ and $E/N$ at the saturation point is
zero within numerical accuracy  for the Hartree-Fock and the 
self-consistent $T-$matrix calculations, and becomes as large as
 $10.7$MeV for the  $G-$matrix approximation. As previously observed 
the use of
 on-shell ({\it i.e.} non self-consistent) $T-$matrix approximation 
\cite{jong}
instead of the $G-$matrix reduces the violation of the Hugenholz-Van Hove
theorem at the saturation point. However, in the case where the 
pairing effect 
is strong (as in this work) the use of the quasi-particle $T-$matrix  
does not cure the violation of the Hugenholz-Van Hove property and 
moreover deforms the results for the binding energy and the effective mass.
We observe a very good fulfillment of the Hugenholz-Van Hove condition 
in the numerical solutions for $\Phi$ derivable approaches with self-energies
self-consistently taken into account, which means for the $T-$matrix 
calculation the use of self-consistent spectral functions in the propagators.

\begin{figure}

\centering
\includegraphics[width=0.48\textwidth]{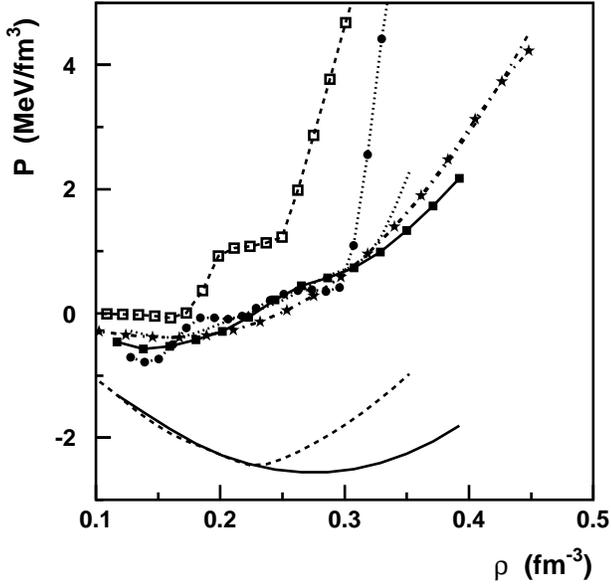}
\caption{Pressure obtained as a derivative of  the binding energy 
(Eq. \ref{press1})
for the
$G-$matrix calculation (solid line), the 
on shell $T-$matrix calculation (dashed line), the full $T-$matrix calculation
(dotted line) and the Hartree-Fock calculation (dashed-dotted line)
as function of density.
The corresponding pressures obtained using Eq. \ref{press2}
 are denoted by the same lines with solid boxes, open boxes,
full circles and stars for the $G-$matrix, on shell $T-$matrix, full $T-$matrix
and Hartree-Fock results.}
\label{pressurefig}
\end{figure}

The pressure can be calculated for a range of densities by two methods 
(Eqs. \ref{press1}, \ref{press2})
 which should be equivalent. However only for the consistent 
approximations we find an approximate equivalence between the 
two formulas, with very good agreement for 
the Hartree-Fock calculation (Fig. \ref{pressurefig}).
 On the other hand  non-consistent
approaches give very different results. In particular the 
point where the pressure equals zero and 
the slope of the pressure versus density comes out differently for the two 
ways of calculating the pressure. The slope of the pressure as function of
 density defines the compression modulus of nuclear matter 
\begin{equation}
{\cal K}=9\frac{\partial P}{\partial \rho}
\end{equation}
which should be positive at the saturation point as a condition of stability.

As expected \cite{fw}  non-conserving approximations give reasonable
 results for the binding energy and not for the Fermi energy.
 Thus, one should use Eq. \ref{press1}
for the calculation of thermodynamic properties
(In particular ${\cal K}=9\frac{\partial}{\partial\rho}\big(\rho^2 
\frac{\partial}{\partial \rho}\big(\frac{E}{N}\big)\big)$). 
The compression modulus for different approximations 
is given in the Table. Its value is similar for different approximations
 using ladder resummation. The values of $\cal K$ obtained are smaller than in 
usual nuclear matter calculations because we reduced the strength of 
the repulsive core.

\section{Conclusion}

We have investigate the thermodynamical consistency of
 different approximations for nuclear matter. The $\Phi$ derivable 
$T-$matrix approximation with off-shell propagators is conserving and 
fulfills these relations. The same is true for the simple Hartree-Fock
approximation. On the other hand the usual $G-$matrix approximation
violates badly the Hugenholz-Van Hove relation for pressure at zero 
temperature. The disagreement is reduced but not cured completely when
using a simplified version of the conserving $T-$matrix approach, 
{\it i.e.} when using the $T-$matrix with on-shell quasi-particle propagators.
The full $T-$matrix  and the $G-$matrix 
calculations give similar results for $E/N$.
The binding energy is a physical result that can be used for 
the calculation of 
the pressure or compression modulus also in the non-conserving 
$G-$matrix approach. 
The same is not true for the Fermi energy which for non-consistent 
approaches is unreliable and leads often to unphysical results if used in
the thermodynamical relations.
We note that the use of the quasi-particle approximation in the $T-$matrix 
resummation can lead to wrong results for the the binding and Fermi energies
if the effect of  pairing is important.
The self-consistent $T-$matrix and the $G-$matrix calculations 
are less sensitive to the fact that we have neglected the superfluid 
transition  for cold nuclear matter.

\begin{acknowledgement}
This work was supported in part by the Polish Committee
 for Scientific Research (KBN)
under Grant 2P03B02019.
\end{acknowledgement}


\end{document}